\documentclass[twocolumn]{aastex631}

\accepted{\today}

\submitjournal{ApJ}

\begin{document}

\title{The Onset of Feedback in Abell 1885: \\ Evidence for Large-Scale Quenching Despite a Young Central AGN }

\author[0000-0001-6699-1300]{Laurel White}
\affiliation{Massachusetts Institute of Technology, 77 Massachusetts Avenue, Cambridge, MA 02139, USA}

\author[0000-0001-5226-8349]{Michael McDonald}
\affiliation{Massachusetts Institute of Technology, 77 Massachusetts Avenue, Cambridge, MA 02139, USA}

\author[0000-0001-5338-4472]{Francesco Ubertosi}
\affiliation{Dipartimento di Fisica e Astronomia, Università di Bologna, via Gobetti 93/2, I-40129 Bologna, Italy}
\affiliation{Istituto Nazionale di Astrofisica (INAF) -- Istituto di
Radioastronomia (IRA), via Gobetti 101, I-40129
Bologna, Italy}


\author[0000-0003-2754-9258]{Massimo Gaspari}
\affiliation{Department of Physics, Informatics and Mathematics, University of Modena and Reggio Emilia, 41125 Modena, Italy}

\author[0000-0001-7271-7340]{Julie Hlavacek-Larrondo}
\affiliation{Département de Physique, Université de Montréal, Succ. Centre-Ville, Montréal, Québec, H3C 3J7, Canada}

\author[0000-0001-5208-649X]{Helen Russell}
\affiliation{School of Physics \& Astronomy, University of Nottingham, University Park, Nottingham NG7 2RD, UK}

\author[0000-0003-3521-3631]{Taweewat Somboonpanyakul}
\affiliation{Department of Physics, Faculty of Science, Chulalongkorn University, 254 Phyathai Road, Patumwan, Bangkok 10330, Thailand}



\begin{abstract}

We present a new 8.5\,ks \emph{Chandra} observation of Abell 1885 (z=0.089), obtained as part of the Cluster Evolution Reference Ensemble At Low-$z$ (CEREAL) survey of $\sim$200 low-$z$ galaxy groups and clusters. These data reveal that Abell 1885 is a strong cool core, with a central cooling time of $0.43_{-0.04}^{+0.12}$ Gyr, and that the central galaxy hosts an X-ray luminous point source at its center (L$_{2-10 \mathrm{~keV}} = 2.3_{-0.7}^{+0.9} \times 10^{42}$ erg/s), indicative of a rapidly accreting supermassive black hole. In the context of the larger CEREAL sample, we
constrain the fraction of clusters at $z\sim0.15$ with X-ray bright (L$_{2-10} > 10^{42}$ erg/s) central AGN to be no more than 4.1\% at 95\% confidence. 
Including radio data from LOFAR, GMRT, ASKAP, and the VLA, spanning 44\,MHz -- 150\,GHz, and optical integral field unit data from SDSS MaNGA, we probe the details of cooling, feeding, and feedback in this system.
These data reveal that cooling of the intracluster medium is highly suppressed on large ($>$10\,kpc) scales despite a central supermassive black hole that is in the early stages of the self-regulation cycle (characterized by rapid accretion, physically small jets, and no large-scale low-frequency radio emission). To reconcile the large-scale quenching with a lack of visible large-scale feedback, we propose that the timescale on which energy is dissipated on large scales ($>$10\,kpc) is significantly longer than the timescale on which black hole feeding operates on small ($\sim$pc) scales. 
%
This interpretation disfavors a model in which the energy is rapidly dissipated (e.g., shocks), which would synchronize the feeding and feedback timescales, and favors a model in which the heating effects of AGN feedback can linger long after the outburst has passed (e.g., turbulent mixing).

\end{abstract}



\section{Introduction}

Decades of X-ray observations of the intracluster medium (ICM) in galaxy clusters---including 25 years with the \textit{Chandra} X-ray Observatory---have revealed a population of cool-core clusters with high densities and low temperatures at their centers. This combination should yield high cooling rates and massive starbursts, yet observations at all wavelengths consistently find lower-than-expected cooling and star formation rates \citep{1987MNRAS.224...75J, 1999MNRAS.306..857C, 2003ApJ...590..207P, 2006PhR...427....1P, 2006ApJ...642..746B, 2008ApJ...681.1035O, 2018ApJ...858...45M}. It is generally accepted that this is due to heating of the ICM from active galactic nuclei (AGN), which accrete cooling gas and inject energy back into the cluster, counteracting cooling and star formation in a process referred to as AGN feedback \citep{2007ARA&A..45..117M, 2012NJPh...14e5023M, 2012ARA&A..50..455F, 2020NatAs...4...10G}. Previous analyses of cluster populations have determined that nearly every cool-core cluster hosts a radio-loud AGN in its brightest cluster galaxy (BCG), strongly implying that AGN feedback is an important component of the cooling cycle \citep{2009ApJ...704.1586S}. However, the specific mechanisms of this process---such as how the energy is imparted into the gas and what the AGN duty cycle looks like---remain poorly understood.

Radio-loud AGN tend to be associated with low ($<<$1\%) rates of accretion onto the black hole \citep{2005MNRAS.363L..91C,2013MNRAS.432..530R}. Alternatively, or sometimes additionally, AGN can operate in quasar (also known as radiative) mode, where intense radiation creates winds that push gas out from the center of the galaxy, and the immediate vicinity of the supermassive black hole. This leads to a decrease in the accretion rate as the AGN transitions to radio mode and launches relativistic jets that mechanically affect the surrounding gas \citep{2013MNRAS.432..530R}. The accretion could be fueled by a steady-state Bondi flow \citep{2013MNRAS.432..530R, 2015MNRAS.451..588R, 2018MNRAS.477.3583R} or a chaotic rain of cold clouds via a process known as chaotic cold accretion (CCA; \cite{2017MNRAS.466..677G}).

\begin{figure}[t!]
\includegraphics[width=\linewidth]{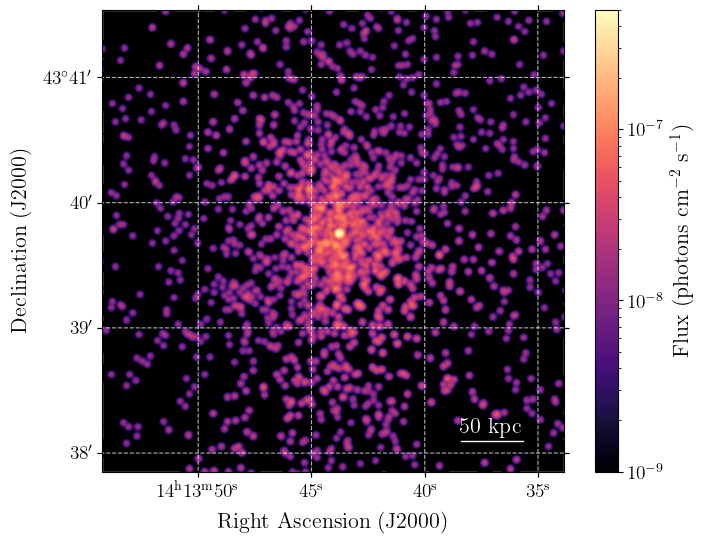}
\caption{An exposure-corrected image of the X-ray flux from Abell 1885 in the 0.5-7.0 keV band as observed with \textit{Chandra}. Standard processing has been applied as described in the text, as well as Gaussian smoothing at the 2$\sigma$ level.}
\label{fig:xray}
\end{figure}

While the vast majority of cool-core clusters are associated with radio-loud AGN, there is a continuous distribution of supermassive black hole (SMBH) accretion rates in these systems with a tail extending towards high Eddington ratios, as demonstrated in \cite{2013MNRAS.432..530R}. There are only five known systems that have high ($>$0.1) Eddington accretion rates leading to central quasars in clusters---Phoenix, H1821, IRAS09104, 3C 186, and 3C 295---but there are also a handful, such as Perseus and PKS\,1353-341, that have X-ray point sources, suggesting a higher-than-average accretion rate compared to other cool cores \citep{2018ApJ...863..122S, 2018ApJ...859...65Y}. \cite{2013MNRAS.431.1638H} suggest that these high accretion rate systems may be more prevalent at high-z, though current samples are fairly biased. A complete census of clusters is necessary to assess the actual frequency of such systems.

In an effort to conduct a fair census of different phenomena in clusters---such as cool cores, central point sources, and mergers---we have assembled a mass- and redshift-limited sample known as the Cluster Evolutionary Reference Ensemble At Low-z (CEREAL) (White et al. (in prep)). CEREAL is comprised of 108 clusters selected using the \textit{Planck} 2nd Sunyaev-Zeldovich Survey \citep{2016A&A...594A..27P} to have redshifts in the range $0.15 < z < 0.25$ and masses in the range $4.5 \times 10^{14}$ M$_{\odot}$ $<$ M$_{500}$ $<$ $10 \times 10^{14}$ M$_{\odot}$. A low-mass extension of the CEREAL sample adds another 63 clusters with redshifts in the range $0.1 < z < 0.2$ and masses in the range $2 \times 10^{14}$ M$_{\odot}$ $<$ M$_{500}$ $<$ $4.5 \times 10^{14}$ M$_{\odot}$. Combined, this yields a sample of 171 massive halos, spanning nearly an order of magnitude in mass at roughly fixed redshift, providing an opportunity to understand the demographics of galaxy groups and clusters in the nearby universe. We have currently observed more than 100 of these clusters, and our observation of Abell 1885 at z=0.089 was the first to demonstrate evidence of a central point source that is visible in the X-ray. Here, we present a multiwavelength assessment of this cluster with the goal of understanding how unique it is compared to the general population.

This paper is structured as follows. In \S2, we present the new \textit{Chandra} observations of Abell 1885, along with the methodology used to reduce these data. We also present the complementary data employed in this work in the optical band (SDSS MANGA) and in the radio band (VLBA, flux density measurements). In \S3, we present the key results of this work, namely the properties of the central X-ray-bright point source, the thermodynamic profiles of the hot ICM, and the radio properties of the central AGN. In \S4, we assess the rarity of systems like Abell 1885 with central X-ray point sources compared to the general population of galaxy clusters and discuss the implications of this specific system with regards to a cyclic AGN feedback cycle. Finally, in \S5, we will summarize our findings and discuss potential avenues for future work.

For this paper, we use a standard $\Lambda$CDM cosmology with H$_0$ = 70 km s$^{-1}$ Mpc$^{-1}$, $\Omega_{\rm{M}}$ = 0.3, and $\Omega_{\Lambda}$ = 0.7. This work employs a list of \textit{Chandra} datasets, obtained by the \textit{Chandra} X-ray Observatory, contained in~\dataset[DOI: 10.25574/cdc.380]{https://doi.org/10.25574/cdc.380}.

\section{Data}

\subsection{Chandra}

We obtain 8.53 ks of ACIS-I data from \textit{Chandra} centered on Abell 1885 as part of a Cycle 25 program to extend the Cluster Evolutionary Reference Ensemble At Low-z (CEREAL) sample to lower-mass clusters. We use the redshift and M$_{500}$ values from the \textit{Planck} 2nd Sunyaev-Zeldovich Survey. We perform standard reprocessing using the CALDB 4.11.6 calibration files. We begin with the CIAO tool \texttt{chandra\_repro}, which applies a charge transfer inefficiency correction, a time-dependent gain adjustment, a gain map, a bad pixel file, pulse height analysis (PHA) randomization, a sub-pixel adjustment, and VFAINT background cleaning, as well as standard grade, status, and good time filters \citep{2006SPIE.6270E..1VF}. We identify point sources using the \texttt{wvdecomp} tool from the ZHTOOLS package, which implements the method described in \cite{1998ApJ...502..558V}, and remove those regions from the analysis. We then use the  CIAO \texttt{dmextract} tool to extract a lightcurve binned on a 259.8 second time scale and the \texttt{deflare} tool to identify and remove data recorded during times in which the lightcurve shows flaring of at least 2.5$\sigma$ significance \citep{2006SPIE.6270E..1VF}. This observation contains no significant flares, so we retain the full 8.53 ks of exposure time. We show an image of the 0.5-7.0 keV emission after these standard processing steps have been completed in Figure \ref{fig:xray}.

\subsubsection{Point Source Spectral Fitting}

We extract a nuclear spectrum from a circular region of radius 1$^{\prime\prime}$ around the central point source. We bin this spectrum in energy to have at least 2 counts per bin. Using XSPEC \citep{1996ASPC..101...17A}, we fit the data in the 0.5-7.0 keV range with a model given by \texttt{phabs*zphabs*pow} to account for absorption from our own galaxy, absorption local to the AGN host galaxy, and AGN powerlaw emission. Note that an estimate of the off-source local background count rate leads us to expect only $\sim$0.014 background counts in this 1$^{\prime\prime}$ region, so we neglect the background in the modeling and fitting process. Similarly, we use an annular region just outside of the central point source region to estimate the number of counts contributed by thermal emission; we expect only $\sim$1.6 counts, so we also neglect this contribution. We fix the Galactic column density to the appropriate value for the sky location (using \citealt{1992ApJS...79...77S}) and the redshift to the known value of 0.089. We fit using the \texttt{cstat} statistic to determine the normalization and slope of the powerlaw as well as the column density of the host galaxy absorption. For the best-fitting model (cstat=8.66, dof=13), we find an X-ray luminosity in the rest frame 2-10 keV range of $L_{\rm{2-10~keV}} = 1.53^{+0.42}_{-0.34}\times10^{42}$\,erg/s, along with a powerlaw slope of $1.31 ^{+1.23} _{-0.47}$ and an intrinsic column density with a $1\sigma$ confidence interval ranging from 0 to $1.62 \times 10^{22}~\mathrm{cm}^{-2}$.

\begin{figure}[t!]
\plotone{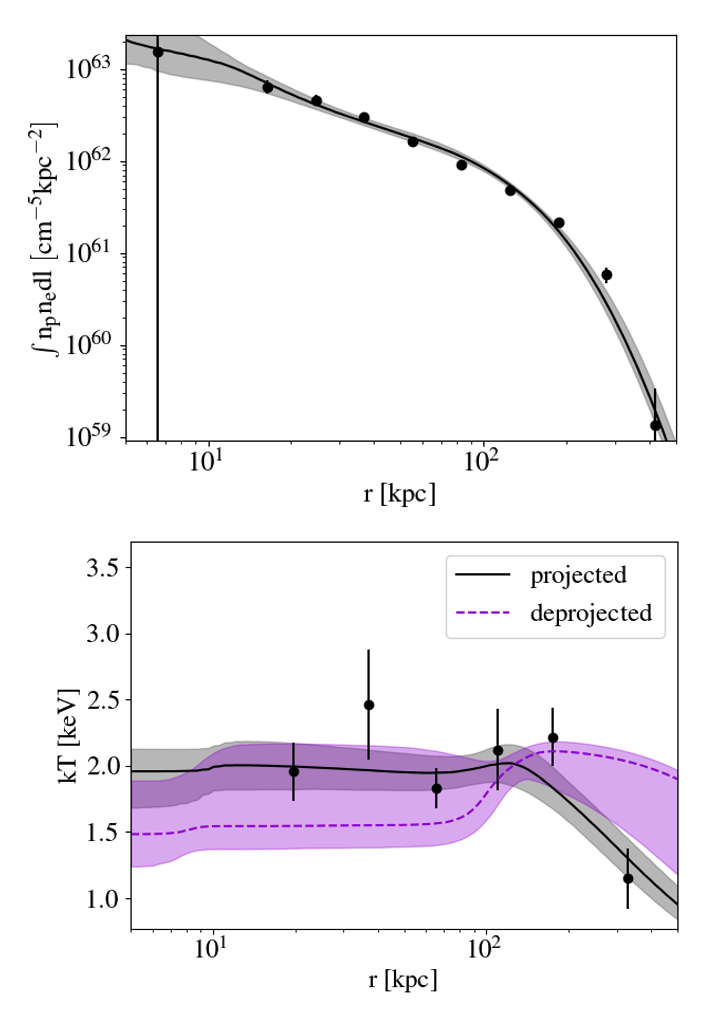}
\caption{Emission measure (top) and kT (bottom) best-fit radial profiles with 1$\sigma$ confidence intervals. The emission measure is proportional to the square of density, and a physically-motivated density profile is integrated to produce the above fit to the observed points. Similarly, the best-fit temperature profile is shown both as it is observed (i.e. projected along the line of sight) as well as unprojected to the true underlying profile. The measured data points with errorbars (used to fit the projected profile) are shown in black.}
\label{fig:thermoprofiles}
\end{figure}

\subsubsection{Global Spectral Fitting}

There are several global properties that can be measured in large apertures and then frozen for subsequent fits with fewer counts, including the background amplitudes, the cluster redshift, and the line-of-sight absorbing column. 
We first extract a single ``full-cluster'' spectrum over $0 < r < 0.5$R$_{500}$ (calculated from the \textit{Planck} M$_{500}$ value). All spectra for this analysis are binned such that each bin contains at least 10 counts.

We use XSPEC to fit a \texttt{phabs*apec} model to the spectrum in the 0.5-7.0 keV region with all parameters free (using the default abundance table from \citealt{1989GeCoA..53..197A}) to obtain the measured values of column density and redshift. We find an offset from the published redshift of dz=0.01 (consistent with the limited energy resolution of ACIS) and no evidence for galactic absorption in the energy range that we are considering. We also find a global temperature of 2.9 keV, metallicity of 0.59 Z$_\odot$, and \texttt{apec} norm of $5.3 \times 10^{-3}$.\footnote{For units, see \url{https://heasarc.gsfc.nasa.gov/xanadu/xspec/manual/node134.html##apec}} In all subsequent fits, we freeze these best-fit values of the redshift and column density.


We also define an annular background region far from the cluster center near R$_{500}$ and extract a new spectrum, which we fit with a \texttt{phabs*pow+apec} model to find the normalization and slope of the \texttt{pow} (unresolved cosmic X-ray background) model and the normalization of the \texttt{apec} (Galactic halo thermal emission) model. The column density is fixed to the value determined above.  We fix the \texttt{apec} temperature, metallicity, and redshift to 0.18 keV, 1.0 Z$_{\odot}$, and 0.0, respectively, to represent the Milky Way halo. The best-fit model is used in all subsequent fits to represent the local astrophysical background, following \cite{2013ApJ...774...23M}, and it also inherently includes the detector background.

\begin{figure}[tb]
\plotone{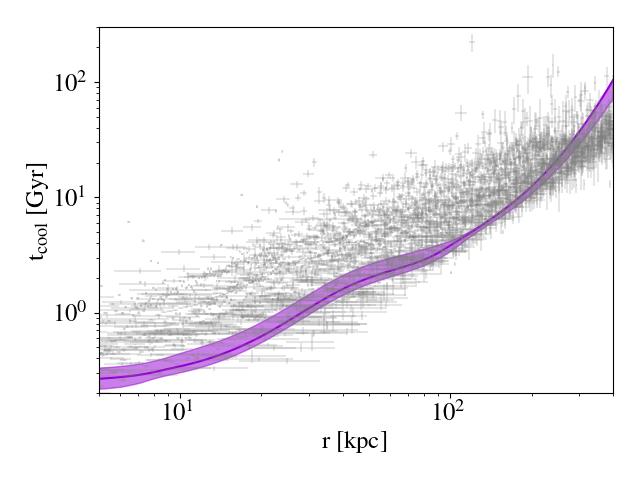}
\caption{Radial cooling time profile, calculated using the best-fit temperature and density profiles, with 1$\sigma$ confidence intervals. The central cooling time is $0.42^{+0.11}_{-0.08}$ Gyr, implying that Abell 1885 is a strong cool core. The background points represent the cooling time profiles of the 91 cool core systems from the ACCEPT sample \citep{2009ApJS..182...12C} for comparison.}
\label{fig:coolprofiles}
\end{figure}

\subsubsection{Thermodynamic Profiles}

Thermodynamic profiles for Abell~1885 were extracted by following a two-stage refinement process, recognizing that some measurements (e.g., metallicity) require significantly more signal to constrain than others (e.g., X-ray emission measure). 
In the first stage of this process, we define 6 annular regions spanning 0.01--0.5R$_{500}$ which contain an average of $\sim$600 counts in each region. The inner boundary was chosen to intentionally exclude the central point source. We fit each spectrum over the energy range 0.5--7.0 keV with a \texttt{const*phabs*apec} model added to the previously determined background model, freezing column density and redshift and fitting only for temperature, metallicity, and normalization of the \texttt{apec} component. This yields 6-bin temperature and metallicity profiles (temperature profile shown in Figure \ref{fig:thermoprofiles}).

The emission measure profile can be constrained with far fewer counts, given knowledge of the temperature and metallicity of the gas as a function of radius. To this end, we fit a simple model for the temperature profile, based on \cite{2006ApJ...640..691V}, to the data:
\begin{equation} \label{temp}
    T(r) = T_0  \frac{(r/r_{\mathrm{cool}})^{a_\mathrm{cool}}+T_{\mathrm{min}}/T_0}{((r/r_{\mathrm{cool}})^{a_\mathrm{cool}}+1)} \frac{1}{[1+(r/r_t)^2]^{0.45}},
\end{equation}
and a simple metallicity profile given by
\begin{equation} \label{metal}
    X(r) = C_1 + \frac{C_2}{\sqrt{2\pi\sigma^2}}e^{-(r-a)^2/2} + C_3 \log_{10}{r}.
\end{equation}

\noindent{}These profiles are used simply to interpolate the temperature and metallicity as a function of radius, for the purpose of estimating an emission measure profile. For this, we extract X-ray spectra in 10 annular regions with radii spanning 0.01--0.5R$_{500}$. We again use a \texttt{const*phabs*apec} model to fit these spectra, but now fitting only for the normalization of the \texttt{apec} component. In these fits, we freeze all other parameters, using interpolations to the metallicity and temperature described above. The resulting emission measure ($\int n_e n_p dV$) profile is modeled with an unprojected density profile from  \cite{2006ApJ...640..691V},
\begin{equation} \label{density}
    \begin{aligned}
        n_p n_e = &n_0^2 \frac{(r/r_c)^{-\alpha}}{(1+(r/r_c)^2)^{3\beta-\alpha/2}} \frac{1}{(1 + (r/r_s)^3)^{\epsilon/3}}\\
        &+ \frac{n_{02}^2}{(1+(r/r_{c2})^2)^{3\beta_2}},
    \end{aligned}
\end{equation}


\noindent{}which is projected along the line of sight and within each radial bin. This allows us to back out the unprojected model for the radial electron density profile. We then use this unprojected density profile to project the temperature model (Equation \ref{temp}) and re-fit the measured temperature points with this projected model, assuming a simple approximation for the weighting of multiple temperature components along the line of sight \citep{2006ApJ...640..691V}. The parameters for the temperature profile fit therefore reflect the true underlying (unprojected) model, rather than the observed (projected) profile. The best-fit emission measure and temperature profiles (both projected and unprojected) are shown in Figure \ref{fig:thermoprofiles}, along with the measured temperature and emission measure data points.




We combine the best-fit unprojected density and temperature profiles to produce a cooling time profile given by
\begin{equation}
    t_\mathrm{cool} = \frac{3(n_e+n_p)kT}{2n_en_p\Lambda(T)},
\end{equation}
where $\Lambda(T)$ is the cooling function for a plasma in collisional ionization equilibrium from \cite{1993ApJS...88..253S}, assuming a metallicity of [Fe/H]=-0.05, since cool cores typically have solar metallicities \citep{2015MNRAS.452.4361K}. The cooling time profile is shown in Figure \ref{fig:coolprofiles}, alongside a sample of 91 cool core clusters from \cite{2009ApJS..182...12C} for comparison. 

\begin{figure*}[t!]
\plotone{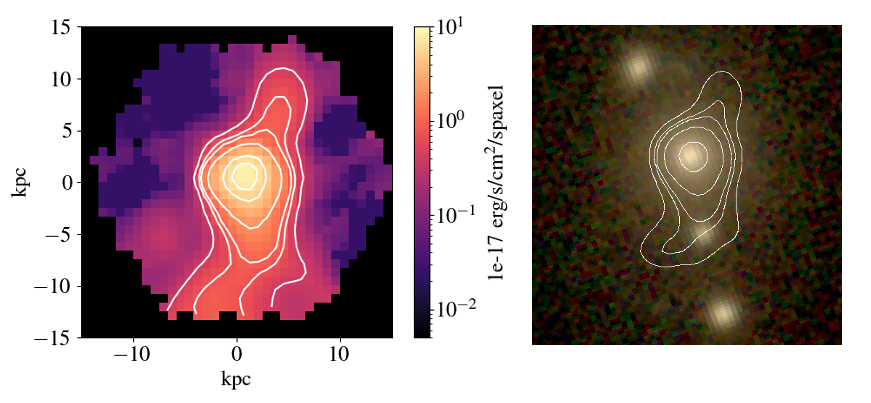}
\caption{\textit{Left}: Map of the integrated H$\alpha$ emission in each pixel of the MANGA IFU cube with contours. There is evidence of star formation extending several kiloparsecs from the center of the cluster. \textit{Right}: 3-band (\textit{g/r/i}) optical image from SDSS data with the same H$\alpha$ contours overlaid. The star formation covers approximately the entire span of the galaxy.}
\label{fig:manga}
\end{figure*}

\subsection{SDSS}

The central galaxy in Abell 1885 is within the SDSS survey footprint and, critically, has available MaNGA \citep{2015ApJ...798....7B} integral field unit (IFU) observations, which we utilize in this work. For each spaxel, we isolate the spectrum in a small wavelength region around the H$\alpha$ and [N\textsc{ii}] emission lines. We fit this wavelength range with a constant value to represent the continuum and three Gaussian emission lines representing the H$\alpha$ line and the N\textsc{ii} doublet. These three lines share a single linewidth and redshift in the fits. Performing this fit on a per-spaxel basis allows us to produce a map of the H$\alpha$ emission, as shown in Figure \ref{fig:manga}. This map reveals a significant amount of warm (10$^4$\,K) ionized gas that is centrally concentrated on the galactic nucleus but extends on scales of $\sim$30\,kpc from end to end. The total H$\alpha$ luminosity in the entire nebula is L$_{H\alpha} = 1.06\times10^{41}$\,erg/s. Note that the emission of the central AGN could contaminate the H$\alpha$ measurement, so this number should be taken as an upper limit.

\subsection{Radio: Surveys and VLBA}
To obtain a comprehensive view of the radio spectral energy distribution (SED) of A1885, we consider flux density measurements from radio surveys and from the literature. The former include the LOFAR Two-metre Sky Survey (LoTSS, 6” resolution, \citealt{2022A&A...659A...1S}) at 144~MHz ($28.2\pm4.0$~mJy), the TIFR GMRT Sky Survey (TGSS, 25” resolution, \citealt{2017A&A...598A..78I}) at 150~MHz ($28.5\pm9.0$~mJy), the Westerbork Northern Sky Survey (WENSS, 54” resolution, \citealt{1997A&AS..124..259R}) at 325~MHz ($29\pm6$~mJy), the Rapid ASKAP Continuum Survey (RACS, 15” resolution,  \citealt{2020PASA...37...48M,2021PASA...38...58H}), at 888~MHz ($39.9\pm2.4$~mJy), and the VLA Sky Survey (VLASS, 2.5” resolution, \citealt{2020PASP..132c5001L}) at 3~GHz ($44\pm1$~mJy). The latter correspond to the flux density measurements reported in \cite{2014PhDT.......338H} and \cite{2015MNRAS.453.1223H,2015MNRAS.453.1201H} at 1.4~GHz, 4.8~GHz, and 150~GHz. Overall, such collection of measurements covers three orders of magnitude in frequency, from 144~MHz to 150~GHz.

We complement the above flux density measurements with Very Long Baseline Array (VLBA) observations at 2.3~GHz (project BE042), 5.0~GHz (project BE063K), and 8.4~GHz (project BE042). We adopted standard reduction techniques in
AIPS to reduce the data,\footnote{\url{http://www.aips.nrao.edu/CookHTML/CookBook.html}.} including a first correction of Earth orientation
parameters and ionospheric delays. Then, we solved for delays and amplitudes, by
applying digital sampling corrections, removing instrumental delays on phases, and calibrating the bandpass.
To correct the time-dependent delays in phases, we fringe-fitted
the data and finally we applied the calibration to
the data. A cycle of phase-only self-calibration was performed on the data. Our final images are characterized by the following spatial beam sizes and rms noise levels: $8.3 \times 2.6$~mas, $\sigma_{\text{rms}} = 0.3$~mJy/beam (2.3~GHz), $4.7 \times 1.7$~mas, $\sigma_{\text{rms}} = 0.05$~mJy/beam (5.0~GHz), and $2.3 \times 0.7$~mas, $\sigma_{\text{rms}} = 0.2$~mJy/beam (8.4~GHz). 


\section{Results}

\subsection{AGN Luminosity, Jet Power, and Accretion Rate}

To convert the measured 2-10 keV rest-frame luminosity of the nucleus into a bolometric luminosity, we consider four different bolometric correction factors from the literature \citep{2004MNRAS.351..169M, 2007ApJ...654..731H, 2012MNRAS.425..623L, 2019MNRAS.488.5185N}. Accounting for the measurement error in the hard X-ray luminosity as well as the range of bolometric corrections in these four publications yields a 1$\sigma$ confidence range for the bolometric luminosity of $4.5 \times 10^{42}-4.2 \times 10^{43}$ erg/s.


In addition, we calculate the AGN jet power, which can be inferred from the radio luminosity following \cite{2010ApJ...720.1066C}. The radio luminosity at 1.4 GHz (see Figure \ref{fig:VLBA}) is found to be P$_{1.4} =  \nu L_{\nu} = 1.2 \times 10^{40}$ erg s$^{-1}$. Folding in the scatter and uncertainty in the relationship between radio luminosity and jet power, we find a jet power of P$_\mathrm{jet} = 102 \pm 66 \times 10^{42}$ erg s$^{-1}$. The total luminosity of the AGN is then the sum of the X-ray and radio luminosities (with the radio luminosity being completely dominated by P$_\mathrm{jet}$), following \cite{2013MNRAS.432..530R}, which is $1.3 \pm 0.7 \times 10^{44}$ erg s$^{-1}$. Assuming $\rm{L}=\epsilon \dot{\rm{M}} \rm{c}^2$ and an accretion efficiency $\epsilon=0.1$, we find a black hole accretion rate (BHAR) of $2.2 \pm 1.2 \times 10^{-2}$\,M$_{\odot}$ yr$^{-1}$. The accretion efficiency for these systems is not well-constrained, but \cite{2021ApJ...908...85M} showed that assuming $\epsilon=0.1$ for both mechanical and radiative powers does not lead to inconsistent results.

The stellar velocity dispersion of the BCG from SDSS spectroscopy \citep{2015ApJ...798....7B} is $230 \pm 8$ km/s. We use the relation from \cite{2013ApJ...764..184M} between the mass of the central black hole, M$_{\mathrm{BH}}$, and the velocity dispersion of the galaxy, $\sigma$, to infer a black hole mass. Combining the uncertainty in the velocity dispersion measurement with the scatter in the M-$\sigma$ relationship yields a black hole mass of 4.7$~\pm~1.1 \times10^8$\,M$_\odot$.

Based on this black hole mass, the Eddington luminosity of this black hole would be $5.8 \pm 1.4 \times 10^{46}$ erg s$^{-1}$, which is $\sim$400 times higher than the combined jet power and bolometric luminosity of the accretion disk, as derived above. This implies that the central SMBH in Abell 1885 is accreting at $0.2 \pm 0.1$\% of the Eddington rate. For comparison, when considering only BCGs with strong evidence of AGN feedback (X-ray cavities), \cite{2013MNRAS.432..530R} find a median accretion rate of 0.07\%, with a hard transition to radiative-dominated power occurring at $>$5\% Eddington. Note that our A1885 data do not show evidence of X-ray cavities, and we suspect there may be none due to the compactness of the radio source. However, we require deeper \textit{Chandra} observations to confirm this, and we defer a more detailed X-ray analysis to Ubertosi et al. (in prep). At accretion rates $<$5\% Eddington, \cite{2013MNRAS.432..530R} find only 6 systems for which the jet and accretion power are comparable (A667, A611, RXCJ0352, Hydra A, Cygnus A, and Zw2089), as is the case for Abell 1885, indicating a rare and perhaps transition state.

\subsection{Cooling and Star Formation}

Generally, evidence for AGN feedback is found in the centers of so-called ``cool core'' clusters, where there is evidence in the X-ray for the ICM to be cooling on timescales much shorter than the age of the Universe. Using the density and temperature profiles derived above (\S2.1, Figure \ref{fig:thermoprofiles}), we estimate a cooling time profile, shown in Figure \ref{fig:coolprofiles}, which yields a central cooling time at $r=10$\,kpc of $0.34^{+0.10}_{-0.04}$ Gyr.  Abell 1885 is consistent with being a strong cool core by most definitions, including that of \cite{2010A&A...513A..37H}, who recommend a cooling time threshold of 1 Gyr to separate strong from moderate cool cores and 10 Gyr to separate non-cool cores from moderate cool cores.

In addition, we can use the cooling time profile to estimate the total amount of cooling in the absence of feedback. We calculate a time-averaged maximal cooling rate by integrating the gas mass interior to the radius at which the cooling time is 7.7 Gyr and then dividing by 7.7 Gyr, following \cite{2018ApJ...858...45M}. This yields a maximal cooling rate of $198\pm30$\,M$_\odot$/yr in an idealized situation with no sources of heating or mixing, where the uncertainty is dominated by the uncertainty in the cooling time at small radii.

We detect a significant amount of extended H$\alpha$ emission on $\sim$30\,kpc scales, as seen in Figure \ref{fig:manga}. Assuming that this gas is photoionized by young stars, we can estimate a total star formation rate based on \cite{1998ARA&A..36..189K}. Given that some of this H$\alpha$ is likely produced by the central AGN, we place an upper limit on the total H$\alpha$-derived star formation rate of $<$0.4 M$_{\odot}$ yr$^{-1}$, which is on the lower end of typical for cool core clusters and is near the maximum level at which stellar mass loss can contribute to ongoing star formation in a BCG \citep{2018ApJ...858...45M}.

Previously, \cite{2008ApJ...681.1035O} reported a star formation rate for Abell 1885 of 5.1 M$_{\odot}$ yr$^{-1}$. This was based on measurements of the 8 and 24 $\mu$m flux from \textit{Spitzer}, which were used to estimate the 15 $\mu$m flux and ultimately the infrared luminosity in \cite{2008ApJS..176...39Q}. However, these shorter wavelengths can also have elevated flux due to dusty AGN -- we note that the BCG was not detected at 70 $\mu$m, with \cite{2008ApJS..176...39Q} placing an upper bound on the emission at this wavelength of 20 mJy. Converting this to a star formation rate according to \cite{2010ApJ...725..677L} yields an estimate of $<0.16$ M$_{\odot}$ yr$^{-1}$, which is consistent with our H$\alpha$-derived star formation rate upper limit and inconsistent with the 24$\mu$m-derived (and likely AGN-contaminated) rate.

\subsection{Radio Properties}

\cite{2014PhDT.......338H} previously reported that the radio SED of A1885 has a convex shape with a peak at around $\nu_{p} = 1.5$~GHz, and that there is no evidence for a significant non-core (that is, extended) component. For the above reasons, A1885 has been classified by \cite{2023A&A...673A..52U} as a candidate ``pre-feedback" galaxy cluster, that is, a system in the AGN triggering phase where the central radio galaxy extends on subkiloparsec scales and is young enough so that no cavities or shocks could have impacted the ICM in the last $\gtrsim10^{8}$~yr.

\begin{figure*}[t!]
\centering
\includegraphics[width=\linewidth]{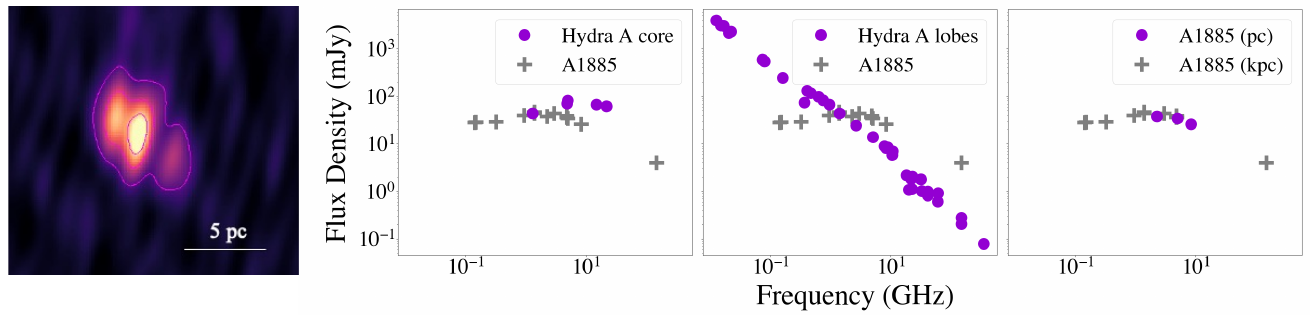}
\caption{\textit{Left Panel}: 8.4 GHz VLBA image with contours. 
\textit{Second panel}: Radio SED of Abell 1885 compared to the small-scale ``core'' component of Hydra A. The Hydra A data have been scaled to match the 1.4 GHx flux density of Abell 1885. \textit{Third panel}: Same as second panel but with the large-scale lobe component of Hydra A rather than the core. \textit{Right panel}: Radio SED of Abell 1885, separated into the parsec- and kiloparsec-scale components. The kpc-scale measurements come the publicly-available radio survey data, while the pc-scale measurements are derived from the VLBA observations. The Abell 1885 emission is completely consistent with the small-scale emission from Hydra A and shows no evidence of an extended component \citep{2014PhDT.......338H}.}
\label{fig:VLBA}
\end{figure*}

From the updated list of radio observations of A1885 (\S 2.3) we can further corroborate the idea that, unlike the vast majority of cool core clusters, there is no evidence for extended, large-scale ($>$kpc) radio emission in the core of Abell 1885. None of the publicly available survey data listed in \S2.3 resolved the radio source at the center of A1885, showing a single, central radio-bright point source. This source remains compact down to the arcsec-scale resolution provided by the VLASS, setting a limit on the radio extent of $R<2.5$\,kpc. Using VLBA data at 8.4~GHz (\S2.3, Figure \ref{fig:VLBA}), we observe extended radio emission on parsec scales, with a linear size of about 7~pc. The source is resolved into a bright central component (the core) and two-sided jets to the east and west of the core. The 2.3~GHz and 5.0~GHz maps (not shown here) are consistent with the pc-scale morphology at 8.4~GHz.

The SED of A1885 can be compared to ``mature-feedback" systems (see \citealt{2023A&A...673A..52U}) with radio emission extended on $>$100 kpc scales (see, for example, \citealt{2023MNRAS.519..767B}). The radio SED shown in Figure \ref{fig:VLBA} reports both the kpc-scale and the pc-scale flux density measurements of A1885 from the data listed in \S2.3. The total flux densities of the pc-scale source are $37.8\pm3.9$~mJy at 2.3~GHz, $34.3\pm3.0$~mJy at 5.0~GHz, and $26.1	2.6$~mJy at 8.4~GHz, which are $\sim100$\% of the kpc-scale flux densities. The continuity between the VLBA data with parsec-scale resolution and the data with kpc-scale resolution suggests that the emission at all frequencies is coming from the smallest scales probed by VLBA \citep{2015MNRAS.453.1223H, 2015MNRAS.453.1201H, 2014PhDT.......338H}. The radio spectrum is consistent with the core emission from, for example, Hydra A \citep{2014PhDT.......338H}, and is inconsistent with the steeply-rising slope expected for radio lobes. The above results confirm the predictions of \cite{2023A&A...673A..52U} that the central AGN of A1885 has recently ``turned on", without evidence for ongoing---or recent---large-scale mechanical feedback. Our X-ray results further indicate that the AGN has awakened following a recent increase in the accretion rate, supporting a more chaotic accretion model (e.g., CCA) rather than a steady flow as predicted by the Bondi accretion model. We will return to this in the discussion in \S4.2.

\begin{figure}[htb!]
\epsscale{1.14}
\plotone{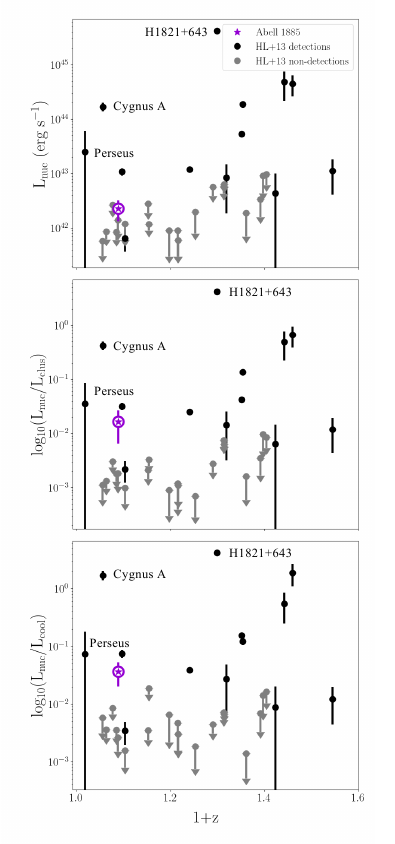}
\caption{Abell 1885 compared to other low-redshift clusters with bright central X-ray point sources from \cite{2013MNRAS.431.1638H}. The top panel shows the hard X-ray (2-10 keV) luminosity of the AGN, L$_\mathrm{nuc}$, as derived from spectral fitting (or 3$\sigma$ upper limits for non-detections). The middle panel shows the ratio of L$_\mathrm{nuc}$ to the overall 0.1-2.4 keV luminosity of the cluster. The bottom panel shows the ratio of L$_\mathrm{nuc}$ to the cooling luminosity, which is defined as the 0.1-2.4 keV luminosity of the cluster out to the radius where t$_\mathrm{cool}$ is equal to the lookback time at z=1 i.e. 7.7 Gyr (for A1885, $\sim$149 kpc).}
\vspace{-0.2in}
\label{fig:rarity}
\end{figure}

\section{Discussion}

\subsection{Rarity of System}

One motivating factor for building out the CEREAL sample is to look for rare systems such as Abell 1885, and to quantify how rare such systems are. After examining data from the 135 CEREAL clusters observed so far, this is the only system harboring a central X-ray-bright point source. We can therefore use binomial statistics to ascertain with 95\% confidence that no more than 4.1\% of clusters at z$\sim$0.15 host an unabsorbed X-ray bright central AGN with $L_{2-10 \mathrm{~keV}} > 10^{42}$ erg/s. While we will refine this estimate once the CEREAL observations are completed, the fact that these 135 clusters represent a nearly random sampling from a mass-complete sample suggest that this is not far from the truth. This preliminary estimate of the fraction of cluster hosting X-ray point sources is consistent with, but less biased than, previous measurements. For instance, \cite{2013MNRAS.431.1638H} finds only 4 systems at $z<0.2$ with detectable X-ray nuclei, which comprise $\sim$2.7\% of the 150 clusters with $z<0.2$ in the Archive of Chandra Cluster Entropy Profile Tables (ACCEPT) catalog \citep{2009ApJS..182...12C}. We expect this number to be biased towards the higher end because of the high luminosities of systems with central X-ray point sources, which will make them more prevalent in flux-limited samples. Note that \cite{2018ApJ...859...65Y} measure a much larger fraction ($>$10\%) of X-ray luminous BCGs in clusters, but they use a significantly larger extraction region for the nucleus. This detection, therefore, places Abell 1885 in a very small minority of systems.

We refer to Figure \ref{fig:rarity} to place Abell 1885 in the context of the handful of other low-redshift clusters with detectable central X-ray point sources as described by \cite{2013MNRAS.431.1638H}. In terms of hard X-ray luminosity, the AGN in Abell 1885 is relatively faint compared to the AGN in, for example, Perseus. However, when we consider the environments of these AGN, Abell 1885 stands out significantly more. When the X-ray luminosity of the central AGN is normalized to either the total cluster luminosity or the cooling luminosity in the core, Abell 1885 is comparable to systems like Perseus, as can be seen by the relative brightness of the AGN and the cluster emission in Figure \ref{fig:xray}. Relative to the host cluster, Abell 1885 has the fourth brightest X-ray AGN at $z<0.2$, sitting about an order of magnitude above other detections and upper limits.

Additionally, this is the lowest luminosity system for which we have been able to calculate the BHAR directly from the X-ray luminosity. For less luminous systems, we have typically had to rely on measurements of the jet power required to create X-ray cavities to back out the BHAR \citep{2021ApJ...908...85M}. We find a ratio of the BHAR to SFR of $\log_{10}(\frac{\mathrm{BHAR}}{\mathrm{SFR}}) = -1.2$, which is consistent with what is found in other BCGs \citep{2021ApJ...908...85M} and elevated compared to what is found in more common, lower mass galaxies (e.g, \citealt{2012ApJ...746..168D, 2015MNRAS.452.3776G, 2015ApJ...808..159X}).

\subsection{Implications for the AGN Feedback Cycle}

In \cite{2023A&A...673A..52U}, it was proposed that Abell 1885 belongs to the class of ``pre-feedback" clusters in which the AGN just turned on and the ICM has not yet been impacted by feedback. This is corroborated by the small, sub-kiloparsec-scale jets coupled with rapid accretion visible in X-rays. Furthermore, survey radio data of the cluster on kiloparsec scales do not show any evidence of extended emission from either current or previous episodes of feedback from the AGN, as described in the previous section.

However, the cooling and star formation rates measured for this system seem to imply a later stage of the cooling/feedback cycle. The observed star formation rate of $<$0.4M$_\odot$/yr is less than 0.25\% of the classically-derived cooling rate, suggesting that $>$99.75\% of the cooling is being offset by some form of heating. The central galaxy in Abell~1885 is even more quenched than the median cool core BCG, which has a star formation rate accounting for 1--2\% of the classical cooling rate \citep{2018ApJ...858...45M}. This system poses the challenge: how can the mechanical feedback just be starting on small ($\sim$pc) scales, yet heating is offset cooling on large ($>10$kpc) scales?

These observations help to refine our understanding of the AGN feedback cycle. They imply that any timescale governing the accretion rate and feedback power output must be decoupled from the timescale on which the energy dissipates into the hot ICM. Specifically, these data argue in favor of a dissipation timescale that is slow compared to the timescale associated with accretion variability (i.e., the AGN duty cycle). This system largely rules out ``instantaneous'' heating mechanisms, such as shocks, in favor of slow, more gentle heating mechanisms, such as turbulent dissipation, as the former would act to couple the cooling timescales on ``micro'' ($\sim$pc) and ``macro'' ($>$10\,kpc) scales. If AGN feedback were purely continuous and lacked episodicity, we would consistently observe a well-defined jet extending on $>$10\,kpc scales, with a declining power-law spectrum in the radio (see e.g., Fig.\ \ref{fig:VLBA}). Instead, the observed state of Abell 1885 suggests that the system is currently in a phase where macro-scale energy dissipation is ongoing while small-scale accretion has resumed, which is typically found in simulations with a self-regulation cycle (e.g., Gaspari et al. 2011, 2012).  Such hydrodynamical simulations show that even in strongly episodic cycling, there is always an irreducible level of turbulence/enstrophy at large scales, as macro-scale ($>$10\,kpc) heating necessarily includes the dissipation of energy from previous AGN episodes, which can remain active via turbulent mixing long after the radio lobes and X-ray cavities have faded (see Wittor et al. 2020).
Thus, we propose that the observed state of Abell~1885 is reflective of the differing timescales on which AGN feeding and feedback operate. Large ($>$10\,kpc) scales evolve slowly as energy disperses via gentle, cascading processes, while on small ($\sim$pc) scales feeding can rapidly resume as small amounts of cold gas either condense or survive past direct-heating episodes, accreting onto the SMBH, boosting the X-ray luminosity of the AGN and leading to the formation of small-scale jets. This framework naturally explains why Abell 1885 can simultaneously exhibit large-scale quenching while also appearing to be at the beginning of a new episode of feedback.

\section{Conclusion}

We have identified Abell 1885 as a system of interest out of a nearly-unbiased sample of 135 low-redshift clusters due to its bright central X-ray point source. X-ray-bright AGN are particularly useful in our mission to understand the particular physics of cooling flows and AGN feedback due to the fact that they signal ongoing accretion and can also serve as ``backlights'' to measure absorption due to line-of-sight cooling.

Through a multiwavelength analysis, we come to the following conclusions:
\begin{enumerate}
    \item Based on the observation of Abell 1885 among a larger, mass-complete, sample of nearby galaxy clusters, we conclude that clusters hosting X-ray-bright central AGN are rare: $\sim$4.1\% of clusters at $z\sim0.15$ host an X-ray bright central AGN with $L_{2-10} > 10^{42}$ erg/s. Upon observing 135 clusters, we find exactly one such system.
    \item We find strong evidence that the AGN in Abell 1885 is quite young. The X-ray-bright nucleus implies ongoing, rapid accretion while the complete lack of large-scale radio emission and very weak low-frequency radio luminosity suggest a lack of large-scale radio lobes. Finally, high angular resolution radio imaging from VLBA show emission on scales of $<$10\,pc, coupled with the lack of large-scale radio emission, suggests that jets have only recently launched.
    
    \item Based on the observation of extended H$\alpha$ emission, we measure a star formation rate in the central galaxy of $\sim$0.4\,M$_{\odot}$ yr$^{-1}$. This is significantly less than the time-averaged maximal cooling rate of $175\pm15$\,M$_\odot$/yr, implying that $>$99.75\% of the radiative cooling of the ICM is being quenched on large scales, \emph{despite the complete lack of evidence for feedback on large scales}.
    
  \item The observation of large-scale suppression of star formation in the presence of a very young AGN implies that the timescale for feedback energy dissipation on large scales is significantly longer than the duty cycle of accretion. These data support a picture in which the mechanical energy from radio-mode feedback is dissipated slowly (e.g., turbulent mixing) and are in tension with processes that heat quickly (e.g., shocks), as these would tend to synchronize the feeding and feedback cycles
  
\end{enumerate}
Ongoing and future work by Ubertosi et al. (in prep) will explore a larger sample of ``young'' feedback systems similar in nature to Abell 1885, both from a radio perspective but also using much deeper X-ray data to search for ghost cavities, providing stronger and more robust constraints on the feedback/cooling duty cycle, which is critically important for understanding how the most massive galaxies in our universe form.

\section*{Acknowledgements}

Support for this work was provided to L.~W. and M.~M. by NASA through Chandra Award Numbers GO2-23117X and GO4-25083 issued by Chandra, which is operated by the Smithsonian Astrophysical Observatory for and on behalf of the National Aeronautics Space Administration under contract NAS803060. 

M.G. acknowledges support from the ERC Consolidator Grant \textit{BlackHoleWeather} (101086804).

%

\vspace{5mm}
\facilities{CXO, SDSS, VLBA}


\software{Astropy \citep{2013A&A...558A..33A, 2018AJ....156..123A, 2022ApJ...935..167A},
          CIAO \citep{2006SPIE.6270E..1VF}
          }



\bibliographystyle{aasjournal}
\bibliography{references}



\end{document}